\documentclass[aps,prb,reprint,a4paper,amsmath,amssymb,amsfonts,noshowpacs]{revtex4-1}
\usepackage{newtxtext,newtxmath}
\usepackage[utf8]{inputenx}
\usepackage{graphicx}
\usepackage[dvipsnames]{xcolor}
\usepackage{soul}
\usepackage[textwidth=17.5cm,textheight=23.5cm,verbose,pdftex]{geometry}
\usepackage[pdftex]{hyperref}

\hypersetup{pdfauthor={Oliver Marquardt}, bookmarksnumbered=true, pdftitle={Modelling the electronic properties of GaAs polytype nanostructures: impact of strain on the conduction band character}, colorlinks, citecolor=blue, linkcolor=blue, urlcolor=blue}

\begin{document}
\graphicspath{{./}}

\title{Modelling the electronic properties of GaAs polytype nanostructures:\\
impact of strain on the conduction band character}

\author{Oliver Marquardt}
\author{Manfred Ramsteiner}
\author{Pierre Corfdir}
\author{Lutz Geelhaar}
\author{Oliver Brandt}
\affiliation{Paul-Drude-Institut für Festkörperelektronik,
             Hausvogteiplatz 5--7,
             10117 Berlin
            }

\begin{abstract}
We study the electronic properties of GaAs nanowires composed of both the zincblende and wurtzite modifications using a ten-band $\mathbf{k}\cdot\mathbf{p}$ model. In the wurtzite phase, two energetically close conduction bands are of importance for the confinement and the energy levels of the electron ground state. These bands form two intersecting potential landscapes for electrons in zincblende/wurtzite nanostructures. The energy difference between the two bands depends sensitively on strain, such that even small strains can reverse the energy ordering of the two bands. This reversal may already be induced by the non-negligible lattice mismatch between the two crystal phases in polytype GaAs nanostructures, a fact that was ignored in previous studies of these structures. We present a systematic study of the influence of intrinsic and extrinsic strain on the electron ground state for both purely zincblende and wurtzite nanowires as well as for polytype superlattices. The coexistence of the two conduction bands and their opposite strain dependence results in complex electronic and optical properties of GaAs polytype nanostructures. In particular, both the energy and the polarization of the lowest intersubband transition depends on the relative fraction of the two crystal phases in the nanowire.  
\end{abstract}

\maketitle

\section{Introduction}

GaAs can be considered as the prototype compound semiconductor material and is used for a wide range of electronic and optoelectronic applications including high electron mobility transistors, solar cells, and infrared laser diodes.~\cite{Wa88,MoJa09} Consequently, the material properties of GaAs have been extensively studied and are known with higher accuracy than for any other compound semiconductor.\cite{Bl82,Ad94,VuMe01} This statement, however, only applies to the equilibrium zincblende (ZB) modification of GaAs, whereas the material properties of the metastable wurtzite (WZ) phase are poorly known. 

This lack of knowledge results from the fact that WZ GaAs cannot be obtained in bulk form or by conventional heteroepitaxy.\cite{McNe05} As a consequence, there has been no need to be concerned with the properties of a metastable phase that escaped investigation in any case. However, this situation has radically changed with the advent of GaAs nanowires (NWs) in which the WZ phase is regularly observed to coexist with the ZB phase in the form of multiple ZB and WZ segments along the NW axis, i.\,e., $\langle 111 \rangle_\text{ZB}$ or $\langle 0001 \rangle_\text{WZ}$.\cite{SoCi05,ZaCo09,HeCo11} The NWs thus constitute polytype heterostructures that are interesting in their own right. However, to unambiguously extract the material properties of bulk WZ GaAs from experiments on these NWs is beset with many difficulties. 

As a consequence, even fundamental properties of WZ GaAs, such as its band gap and the nature of the lowest conduction band (CB), are still controversially discussed.\cite{DePr10,BePa12,GrCo13,BeBe13} While the ZB phase is characterized by a single CB of $\Gamma_{6c}$ symmetry with a light effective mass, two energetically close CBs exist in the WZ phase: the $\Gamma_{7c}$ band, the equivalent of $\Gamma_{6c}$ of the ZB phase with a comparably light effective mass, and the $\Gamma_{8c}$ band, which has no equivalent in the ZB phase, but originates from folding the L valley of the ZB band structure to the center of the Brillouin zone and thus exhibits a heavy and anisotropic effective mass.\cite{TrKi99,BePa12,BeBe13} To our knowledge, all available studies agree that the energy difference between the two CBs in the WZ phase is small ($<0.1$~eV), but differ concerning the ordering of the two bands, namely, whether the $\Gamma_{8c}$ band is energetically below the $\Gamma_{7c}$ band or vice versa.\cite{DePr10,HeCo11,MuNa94,ChLa11,BePa12}

The study of Cheiwchanchamnangij and Lambrecht\cite{ChLa11} has shown that the ordering of these two bands depends sensitively on strain. In particular, for a uniaxial strain parallel to the NW axis, the deformation potentials of the $\Gamma_{7c}$ and $\Gamma_{8c}$ bands were found to be of opposite signs such that the bands cross for small uniaxial strains $\epsilon_{zz}$, with the exact magnitude depending on the equilibrium lattice constants used for the calculation. This theoretical result was experimentally confirmed by Signorello et al.,\cite{SiLo14} who performed experiments on single NWs to which an external uniaxial strain was applied. Signorello et al.\cite{SiLo14} observed the $\Gamma_{7c}$/$\Gamma_{8c}$ crossover at $\epsilon_{zz} = -0.14$\%. We note that a strain of this magnitude may also be introduced unintentionally upon dispersal of the NWs on a substrate.\cite{CoFe15} 

In addition to these extrinsic sources of strain, an intrinsic source exists that has so far been ignored in studies of the electronic structure of GaAs polytype NWs: the non-negligible lattice mismatch between ZB and WZ GaAs. High-resolution x-ray diffraction experiments demonstrate that the in-plane lattice constant $a$ of WZ GaAs is smaller than the equivalent interatomic distance on the (111) plane of the ZB phase by $-(0.27 \pm 0.05)$\%.\cite{McNe05,TcHa06,BiBr12,JoYa15}  Considering the sensitivity of the band structure of WZ GaAs to strain of this magnitude, it is obviously essential to take into account this lattice mismatch for the interpretation of experiments performed on polytypic GaAs NWs. As a consequence, it is imperative for any such interpretation to rely on a model that includes both the $\Gamma_{7c}$ and $\Gamma_{8c}$ bands in WZ GaAs explicitely.

In the present work, we employ and evaluate a ten-band $\mathbf{k}\cdot\mathbf{p}$ model suitable to describe polytype heterostructures represented by two intersecting potentials formed by the $\Gamma_{6c}$ (ZB) and $\Gamma_{7c}$ (WZ) bands as well as by the $\Gamma_{8c}$ (WZ) band which has no equivalent in the ZB phase. The model treats the $\Gamma_{7c}$ and $\Gamma_{8c}$ bands on an equal footing and thus allows us to take into account strain from both the lattice mismatch between the ZB and the WZ phase as well as from external influences. Parameters for (111)-oriented ZB systems are transformed to their respective WZ counterparts such that both crystal phases can be described within the same Hamiltonian. We compute the electronic properties of pure ZB and WZ GaAs NWs as well as of polytypic GaAs NW heterostructures. We show that strain-induced modifications of the two CBs in the WZ phase have a decisive influence on both the character and the confinement of electrons in polytype GaAs heterostructures.

\section{Formalism and parameters}

Our simulations employ a $\mathbf{k}\cdot\mathbf{p}$ Hamiltonian based on the eight-band model for strained WZ semiconductors developed by Chuang and Chang,~\cite{ChCh96} expanded to ten bands with the parabolic $\Gamma_{8c}$ band under the influence of strain.~\cite{ChLa11} . 
This simple approach captures the fundamental feature of the potentials formed by two uncoupled, coexisting CBs in the WZ phase. 

All parameters employed for the calculations are compiled in Table~\ref{tab:params} and were taken from Ref.~\onlinecite{ChLa11} unless indicated otherwise. We have chosen the lattice constants computed within the local density approximation (LDA), since these values are much closer to the experimentally obtained lattice constants\cite{McNe05,TcHa06,BiBr12} than the ones obtained via the generalized gradient approximation (GGA).\cite{ChLa11} As a result, $E(\Gamma_{8c}) < E(\Gamma_{7c})$ at zero strain, contrary to the ordering reported in Ref.~\onlinecite{SiLo14} in which the GGA values were used. This difference reflects the present uncertainty in parameters. In any case, the energy difference between the two bands is small, and the bands cross for uniaxial strains of the same magnitude (but opposite signs). 

The notation of the deformation potentials follows the one of Ref.~\onlinecite{SiLo14}. Lattice, elastic, and piezoelectric constants of the ZB crystal along the $\langle$111$\rangle$ direction and the WZ phase were obtained from the respective ZB values and transformed via the relations given in Ref.~\onlinecite{ScCa11}. As there is no equivalent of the WZ $\Gamma_{8c}$ band in the ZB phase, we assigned a barrier of 1.5~eV to it, which is approximately the energy separation between the $\Gamma_{6c}$ and the next higher CB in the ZB phase. We have assigned the same electron effective masses as in the WZ phase for this band in the ZB segment, since the employed ten-band model requires the consistent treatment of the $\Gamma_{8c}$ band in both crystal phases. The Hamiltonian (see Appendix) was implemented within the generalized multiband $\mathbf{k}\cdot\mathbf{p}$ module of the S/PHI/nX software library.~\cite{BoFr11,MaBo14}

\begin{figure}[t]
\includegraphics[width=\columnwidth]{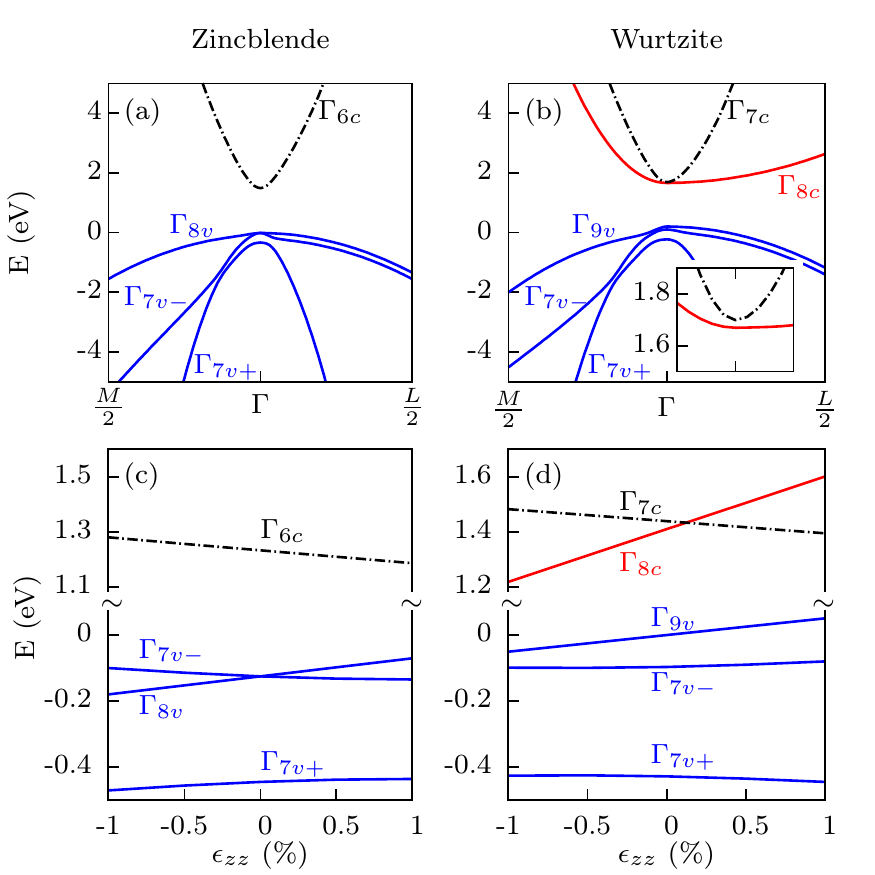}
\caption{(Color online) Band structures of (a) the ZB and (b) the WZ phases of GaAs computed using the parameters given in Table \ref{tab:params}. VBs are depicted in blue, dashed black lines represent the $\Gamma_{6c}$ and $\Gamma_{7c}$ CBs, and the solid red line is the $\Gamma_{8c}$ CB in the WZ phase. (c) ZB and (d) WZ CB and VB energies at the $\Gamma$ point as a function of an external uniaxial strain $\epsilon_{zz}$.}
\label{fig:bs}
\end{figure}

Figure~\ref{fig:bs} shows the bulk bandstructure as well as the response of the band edges at the $\Gamma$ point to an external uniaxial strain $\epsilon_{zz}$ obtained with the parameters listed in Table \ref{tab:params} for both the ZB and the WZ modifications of GaAs. The familiar band structure of ZB GaAs in Fig.~\ref{fig:bs}(a) is different from the band structure of WZ GaAs displayed in Fig.~\ref{fig:bs}(b) not only for the valence bands (VBs), but particularly for the CBs. The energy splitting of the two CBs close to the $\Gamma$ point is visualized in the inset of Fig.~\ref{fig:bs}(b). 

Figures~\ref{fig:bs}(c) and \ref{fig:bs}(d) illustrate the influence of an external uniaxial strain on the $\Gamma$ point CB and VB energies obtained with the parameters listed in Table \ref{tab:params}. For the ZB phase, the VBs are degenerate at zero strain and split at any finite  strain value. For the WZ phase, the VBs are split already at zero strain, and their order does not change within the intervals of strains considered here. However, the character of the lowest CB changes from $\Gamma_{8c}$ to $\Gamma_{7c}$ at $\epsilon_{zz} = 0.12$\%. This change has important consequences: not only does the energy of the optical transition change, but also the oscillator strength.~\cite{BePa12}
 
\begin{table}[t]
\caption{Material parameters for $\langle 0001 \rangle_\text{WZ}$ and $\langle 111 \rangle_\text{ZB} $ GaAs employed within this work. Listed are lattice and elastic constants, piezoelectric constants, spontaneous polarization and the dielectric constant, band gaps, band splittings, and Kane matrix elements, CB effective masses and VB Luttinger-like parameters A$_i$, and band edge deformation potentials. If not indicated otherwise, all parameters are taken from Ref.~\onlinecite{ChLa11}. Values in parentheses were obtained via the cubic approximation or a transformation to translate ZB parameters to the respective WZ ones.}
\label{tab:params}
\begin{ruledtabular}
\begin{tabular}{c|cc}
Parameter & Wurtzite~~ & ~~Zincblende\\
\hline
$a$ (\AA) & 3.955 & (3.9697)\\
$c$ (\AA) & 6.526 & (6.4825)\\
$C_{11}$ (GPa) & \multicolumn{2}{c}{(149.35)$^\text{a}$}\\
$C_{12}$ (GPa) & \multicolumn{2}{c}{(47.52)$^\text{a}$}\\
$C_{33}$ (GPa) & \multicolumn{2}{c}{(158.43)$^\text{a}$}\\
$C_{44}$ (GPa) & \multicolumn{2}{c}{(50.92)$^\text{a}$}\\
\hline
$e_{31}$ (C/m$^2$) & \multicolumn{2}{c}{(0.1328)$^\text{b}$}\\
$e_{33}$ (C/m$^2$) & \multicolumn{2}{c}{($-0.2656$)$^\text{b}$}\\
$P_\mathrm{sp}$ (C/m$^2$) & $-0.0023^\text{c}$ & 0\\
$\kappa_r$ & \multicolumn{2}{c}{13.18$^\text{d}$}\\
\hline
$E_G$ (eV) & 1.554 & 1.503\\
$E(\Gamma_{8c}) - E(\Gamma_{7c})$ (eV) & $-0.029$ & ---\\
VB offset (eV) & 0 & $-0.117^\text{e}$\\
$\Delta_\mathrm{cr}$ (eV) & 0.180 & 0\\
$\Delta_\mathrm{so}$ (eV) & 0.345 & 0.320\\
$E_{P,\parallel}$ (eV) & 28.9 & 28.0$^\text{f}$\\
$E_{P,\perp}$ (eV) & 18.8 & 28.0$^\text{f}$\\
\hline
$m_{\Gamma_{8c},\parallel}$ ($m_0$) & 1.060 & ---\\
$m_{\Gamma_{8c},\perp}$ ($m_0$) & 0.107 & ---\\
$m_{\Gamma_{7c},\parallel}$  ($m_0$) & 0.060 & 0.069\\
$m_{\Gamma_{7c},\perp}$  ($m_0$) & 0.075 & 0.069\\
$A_1$ & $-$18.39 & $-19.3$\\
$A_2$ & $-$1.87 & $-1.4$\\
$A_3$ & 17.05 & 18.0\\
$A_4$ & $-$6.26 & $-9.0$\\
$A_5$ & $-$6.83 & $-8.1$\\
$A_6$ & $-$7.27 & $-10.1$\\
\hline
$\Xi_{d,u}$ (eV) & 21.0 & 0\\
$\Xi_{d,h} - \Xi_{b,h}$ (eV) & 5.16 & 0\\
$\Xi_{b,h} - D_1 -2 D_2$ (eV) & $-8.25$ & $-8.44$\\
$D_3$ (eV) & 7.68 & (8.314)$^\text{a}$\\
$D_4$ (eV) & 7.68 & ($-4.157$)$^\text{a}$\\
\end{tabular}
\end{ruledtabular}
$^\text{a}$ Ref.~\onlinecite{VuMe01}, $^\text{b}$ Ref.~\onlinecite{BeZu06}, $^\text{c}$ Ref.~\onlinecite{ClSe16}, $^\text{d}$ Ref.~\onlinecite{Ad85}, $^\text{e}$ Ref.~\onlinecite{BePa12}, $^\text{f}$ Ref.~\onlinecite{ScWi07}.
\end{table}

\section{Pure zincblende and wurtzite nanowires}
\begin{figure}[t]
\includegraphics[width=\columnwidth]{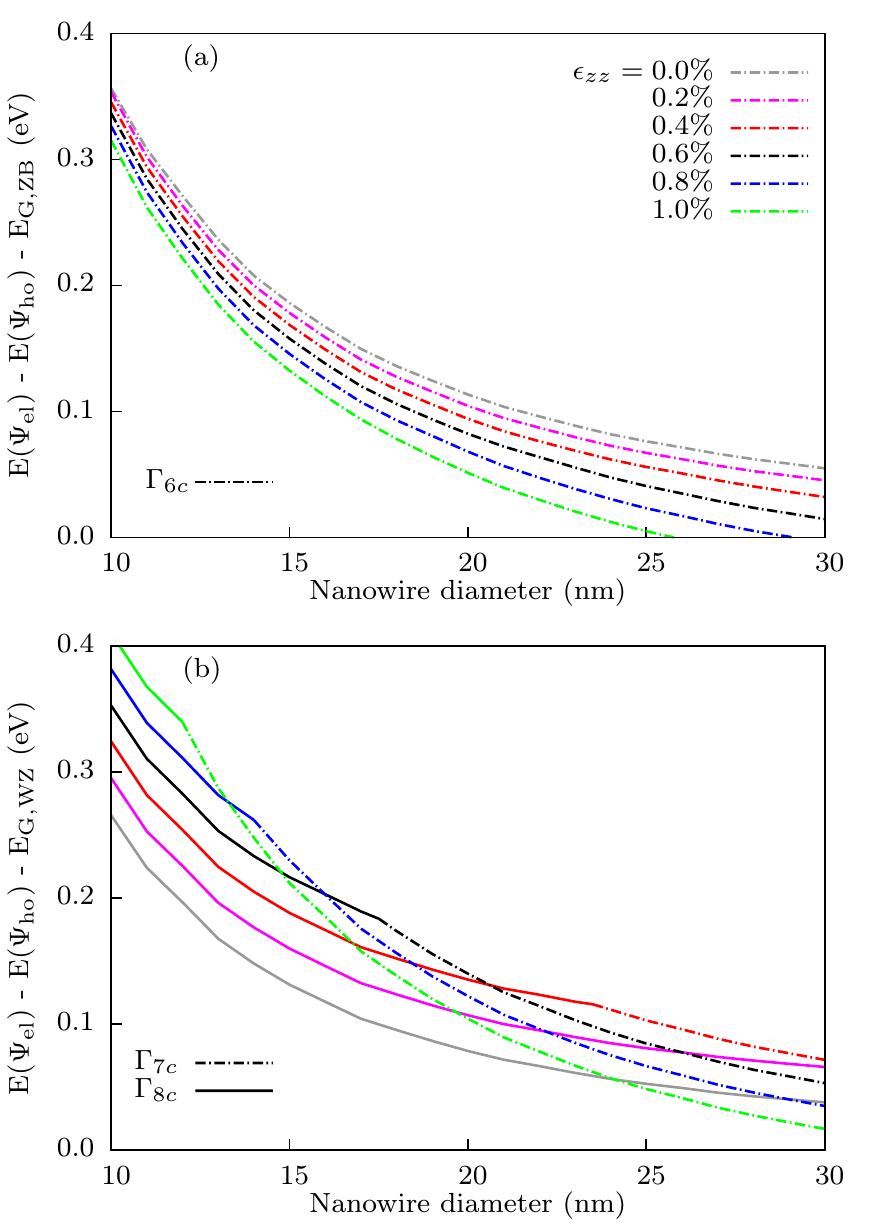}
\caption{(Color online) Energy difference between the electron and hole ground states as a function of the diameter of a purely ZB (a) and a purely WZ (b) GaAs NW relative to the respective unstrained band gap for different external uniaxial strains $\epsilon_{zz}$. Solid lines indicate a $\Gamma_{8c}$ character of the electron state involved, dash-dotted lines indicate a $\Gamma_{6c}$ (a) or $\Gamma_{7c}$ (b) character. Note that the energy of electrons with $\Gamma_{6c}$ and $\Gamma_{7c}$ character decreases with increasing strain whereas electrons with a $\Gamma_{8c}$ character show an opposite behavior.}
\label{fig:NW}
\end{figure}

We start with a discussion of the electronic properties of pure ZB and WZ NWs under the influence of strain and radial confinement. Figure~\ref{fig:NW} shows the energy difference between electron and hole ground states relative to the band gap of the corresponding phase as a function of the diameter of NWs that are subject to an uniaxial strain $\epsilon_{zz}$ of up to 1\%. For an unstrained ZB NW [cf. Fig.~\ref{fig:NW}(a)], the energy decreases with increasing diameter due to a reduced radial confinement, and converges towards the unstrained ZB band gap. Note that dielectric confinement~\cite{ZeCo16} is not considered in this model. For finite tensile strain, the energy is reduced for all diameters.  The electron state has in all cases a $\Gamma_{6c}$ character, as this band is energetically well separated from any other band. The hole ground state is subject to strong band mixing for all NW diameters and strains considered. The character of the hole state thus changes continuously such that no abrupt change of the hole energy is observed. The contribution of the light hole ($\Gamma_{7v-}$) decreases with decreasing diameter and larger strain $\epsilon_{zz}$.

The situation changes entirely when considering a pure WZ NW [Fig.~\ref{fig:NW}(b)]. For the parameter set employed in the present work, the energetically lower band for the unstrained NW is the $\Gamma_{8c}$ band, which exhibits a heavier effective electron mass as compared to the $\Gamma_{7c}$ band. Hence,  the electron ground state is of $\Gamma_{8c}$ character regardless of the NW diameter. 
We furthermore consider the two CBs to be uncoupled, as shown for bulk WZ GaAs~\cite{BeBe13} so that no band mixing occurs. Under the influence of tensile uniaxial strain, the $\Gamma_{7c}$ band is lowered energetically and crosses the $\Gamma_{8c}$ band for $\epsilon_{zz} = 0.12\%$ [cf. Fig.~\ref{fig:NW}(d)]. For larger strains, the electron ground state is of $\Gamma_{8c}$ character up to a certain diameter due to the large effective mass of this band. For larger diameters, the ground state changes its character to $\Gamma_{7c}$ since the influence of quantum confinement diminishes.
Since the two bands are not electronically coupled, this change of character is abrupt, in marked contrast to the behavior known from VB states in ZB GaAs NWs. Hence, WZ GaAs NWs of slightly different diameter or experiencing slightly different strain may exhibit drastically different optical properties in terms of polarization selection rules and oscillator strength. The coexistence of the $\Gamma_{7c}$ and $\Gamma_{8c}$ bands has thus important consequences for the interpretation of experimental results obtained from single NWs.

\begin{figure}[t]
\includegraphics[width=\columnwidth]{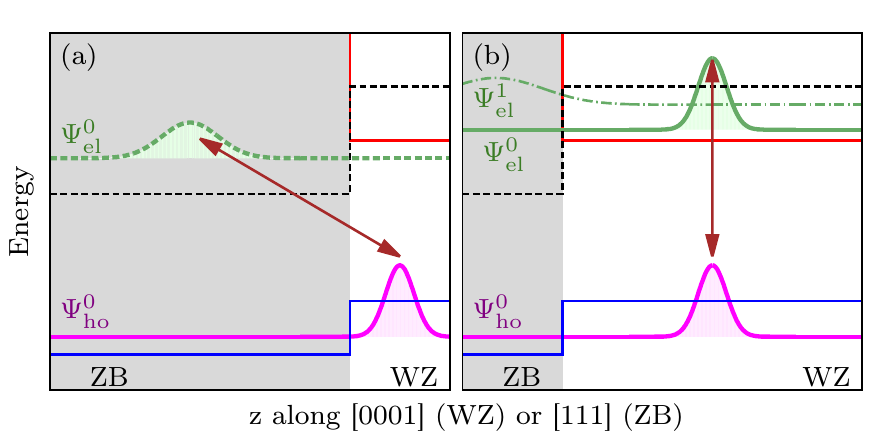}\\
\caption{(Color online) Schematic representation of the two types of optical transitions in polytype NWs. The electrons ground state (green) may be located either in (a) the ZB segment due to the potential offset between the $\Gamma_{6c}$ (ZB) and $\Gamma_{7c}$ (WZ) band (dashed black line) for thick ZB and thin WZ segments or in (b) the WZ segment due to the $\Gamma_{8c}$ band offset (solid red line) for thick WZ segments, while the hole ground state  (purple) always resides in the WZ segment due to the $\Gamma_{8v}$/$\Gamma_{9v}$ potential offset (solid blue line). The corresponding transitions are thus either spatially (a) indirect or (b) direct. The dash-dotted green line in (b) indicates an excited state confined in the ZB segment.}
\label{fig:reco}
\end{figure}

\section{Polytype superlattices}
In this section, we address the electronic properties of WZ/ZB polytype heterostructures as computed in the framework of our ten-band $\mathbf{k}\cdot\mathbf{p}$ model. Since we are interested here in the influence of axial confinement, we restrict the following discussion to NW diameters for which radial confinement can be safely neglected, i.\,e., to diameters larger than 50~nm. Assuming further that other radial contributions to the potential landscape, such as surface potentials induced by Fermi level pinning, can also be excluded, we may approximate GaAs NWs consisting of WZ and ZB segments by a planar polytype heterostructure.

\subsection{Spatially direct and indirect transitions}

It is generally accepted that WZ/ZB heterostructures from III-V semiconductors represent type II heterostructures with the CB minimum in the ZB phase and the VB maximum in the WZ phase.\cite{MuNa94} Consequently, electrons and holes are expected to be spatially separated in these structures. This view, however, is too simplistic in that it neglects the coexistence of two CB in the WZ phase. In fact, the $\Gamma_{6c,7c}$ and the $\Gamma_{8c}$ bands  form two intersecting but not interacting potentials for electrons.
Figure \ref{fig:reco} illustrates that, depending on the length of the segments, both spatially indirect and direct optical transitions are possible in a WZ/ZB heterostructure. In Fig.~\ref{fig:reco}(a), the electron ground state is located in the comparatively long ZB segment due to the potential offset between the $\Gamma_{6c}$ band in the ZB phase and the equivalent $\Gamma_{7c}$ band in the WZ phase [cf. Tab.~\ref{tab:params}]. Since the hole ground state always resides in the WZ segment due to the $\Gamma_{8v}$/$\Gamma_{9v}$ potential offset between the ZB and the WZ phase, the optical transitions are spatially indirect in this case. The situation may change for thin ZB segments as shown in Fig.~\ref{fig:reco}(b). Here, the quantized state in the ZB segment is at an energy higher than the $\Gamma_{8c}$ band in the WZ segment. This band has no equivalent in the ZB segment, which thus represents a high energy barrier for an electron in the WZ segment. For thin ZB segments, the electron ground state is thus confined in the potential well formed by the $\Gamma_{8c}$ band in the WZ segment. Spatially direct transitions between these electrons are allowed with holes in the $\Gamma_{9v}$ VB for a polarization perpendicular to the $\langle$0001$\rangle$ direction with a small, but nonzero dipole matrix element. The green dash-dotted line in Fig.~\ref{fig:reco} (b) indicates the first electron state that is confined in the ZB segment, which is energetically above the ground state confined in the WZ segment.

\begin{figure}[t]
\includegraphics[width=\columnwidth]{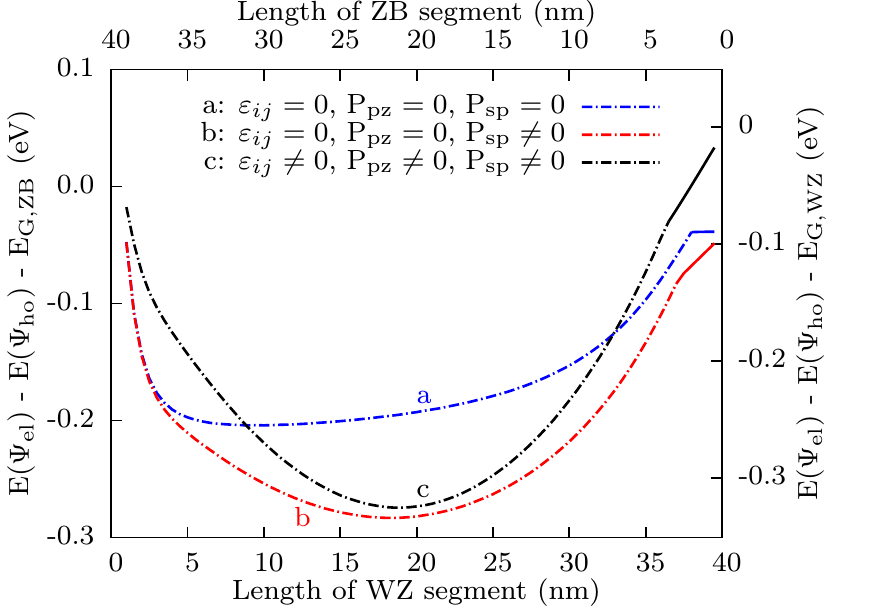}
\caption{(Color online) Energy difference between the electron and hole ground states relative to the ZB (left axis) or the WZ (right axis) band gap for polytype superlattices with different length ratios between the WZ and the ZB segments. The calculations apply to the case of zero external strain and either neglect both the lattice mismatch and thus the intrinsic strain $\varepsilon_{ij}$ and the built-in potentials (blue), or take into account only the spontaneous polarization (red), or include intrinsic strain as well as spontaneous and piezoelectric polarization (black). Solid lines indicate a $\Gamma_{8c}$ character of $\Psi_\mathrm{el}$, whereas dash-dotted lines indicate a $\Gamma_{6c}$ or $\Gamma_{7c}$ character. The length of the super cell is 40~nm.}
\label{fig:contribs}
\end{figure}

\subsection{Intrinsic strain and polarization}
The above qualitative considerations show that it is essential to treat both CBs in the WZ phase on an equal footing. For quantitative results, it is important to note that the electronic properties of ZB and WZ segments in GaAs NWs are modified by strain as well as spontaneous and piezoelectric polarization potentials, $P_\text{sp}$ and $P_\text{pz}$, respectively. The in-plane lattice constants of WZ and $\langle 111 \rangle$-oriented ZB crystals differ by about 0.3\%. Polytype NWs will adopt an average lattice constant that depends on the overall fraction of ZB and WZ segments. These segments are thus under compressive and tensile biaxial strain $\varepsilon_{ij}$ with $i,j=x,y,z$, respectively, which in turn induce a corresponding piezoelectric polarization. In addition, WZ GaAs exhibits a spontaneous polarization of $P_\text{sp}= -2.3 \times 10 ^{-3}$~C/m$^2$ along the $\langle$0001$\rangle$ direction.~\cite{ClSe16} The total polarization discontinuity at the ZB/WZ interfaces gives rise to a polarization potential in polytype NWs composed of ZB and WZ segments. For the following calculations, we consider a superlattice consisting of a ZB and a WZ segment with a total length of 40~nm, and individual lengths between 1 and 39~nm.

\begin{figure}[t]
\includegraphics[width=\columnwidth]{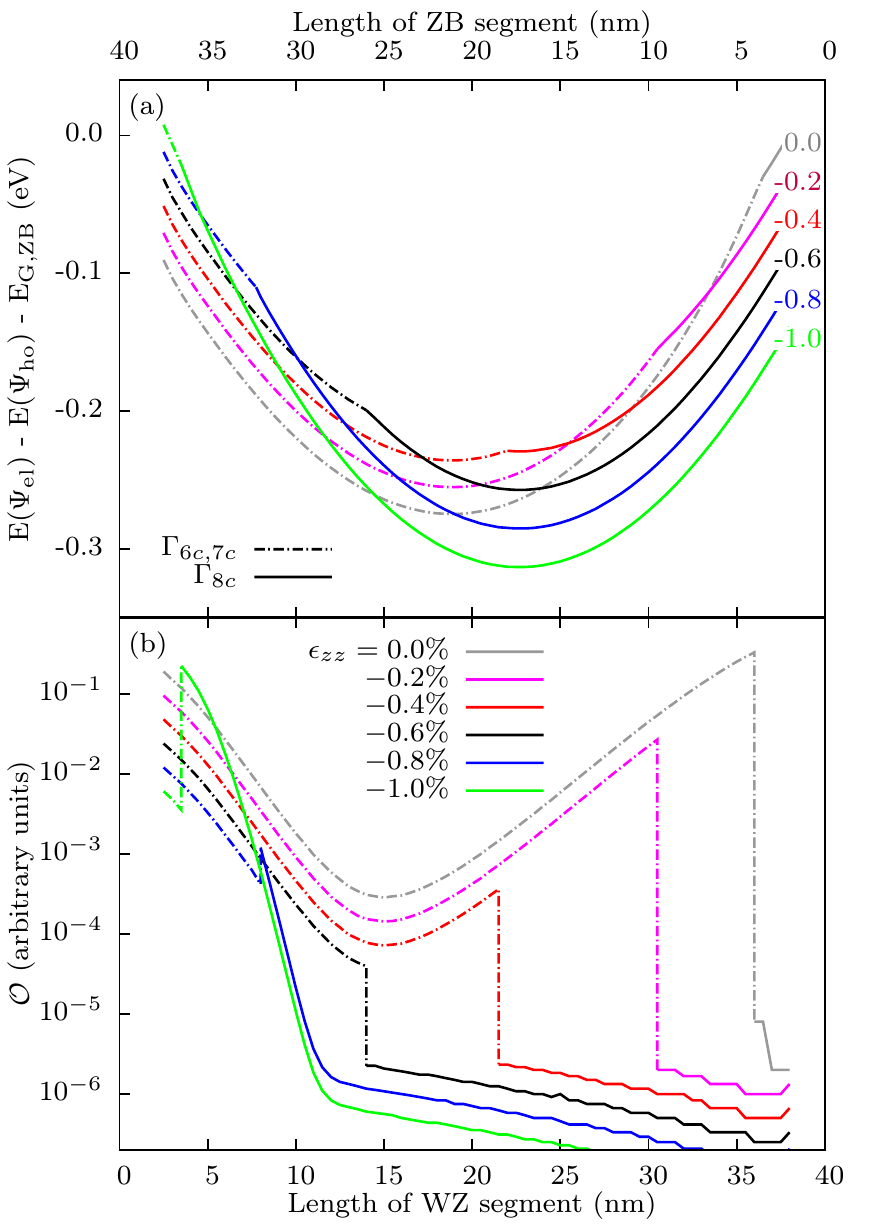}
\caption{(Color online) (a) Energy difference between the electron and hole ground state relative to the ZB band gap as a function of the length of the WZ segment for different values of the external uniaxial strain $\epsilon_{zz}$. Solid lines indicate a $\Gamma_{8c}$ character of the electron wavefunction $\Psi_\mathrm{el}$, whereas dash-dotted lines indicate a $\Gamma_{6c}$ or $\Gamma_{7c}$ character. (b) Electron-hole overlap $\mathcal{O}$ as a function of the length of the WZ segment for different values for the external strain. For the sake of visibility, a constant shift has been added to the individual curves of the overlap.}
\label{fig:SL}
\end{figure}

We first evaluate the influence of internal strain and built-in electric fields on the electronic properties of this WZ/ZB superlattice in the absence of additional external strain. Figure~\ref{fig:contribs} shows the energy difference between electron and hole ground states relative to the band gaps of unstrained ZB and WZ GaAs as a function of the length of the WZ segment. The intrinsic biaxial strain $\varepsilon_{ij}$ within the segments was computed assuming that the equilibrium in-plane lattice constant is given by an average of the ZB and WZ lattice constants weighted by the respective segment length.~\cite{Va89} If both the lattice mismatch and the polarization potentials are neglected [cf. curve a in Fig.~\ref{fig:contribs}], the energy difference between the electron and the hole ground states first drops abruptly due to decreasing hole confinement in the WZ segment, reaches a minimum at a length of 7~nm, and increases for longer WZ segments due to the increasing electron confinement in the ZB segment. The electron remains confined in the ZB segment with a $\Gamma_{6c,7c}$ character (dash-dotted line) up to a WZ segment length of 36~nm. For even longer segments, the electron ground state becomes confined in the WZ segment and its character changes to $\Gamma_{8c}$ (solid line). For all segment lengths, the energy difference between electron and hole with respect to the ZB band gap remains negative, i.\,e., the energy of optical transitions would be below the ZB band gap due to the VB offset between ZB and WZ GaAs.

When we include spontaneous polarization in our simulations, as shown in curve b in Fig.~\ref{fig:contribs}, the overall energy redshifts becomes larger with increasing length of the WZ segment. At the minimum of the curve at a length of 20~nm, the energy shift amounts to 90~meV as compared to curve a. Considering, in addition, the lattice mismatch and the resulting biaxial strain and piezoelectric polarization potentials [cf. curve c], significant differences are observed with respect to curve b both for short and long WZ segments. In particular, for WZ segments longer than 36~nm, the energy difference between the $\Gamma_{8c}$ electron and the $\Gamma_{9v}$ hole states exceeds the band gap of ZB GaAs. Note, however, that we never reach or even exceed the band gap of WZ GaAs, which is a consequence of the presence of internal electrostatic fields in the heterostructure.

\begin{figure}[t]
\includegraphics[width=\columnwidth]{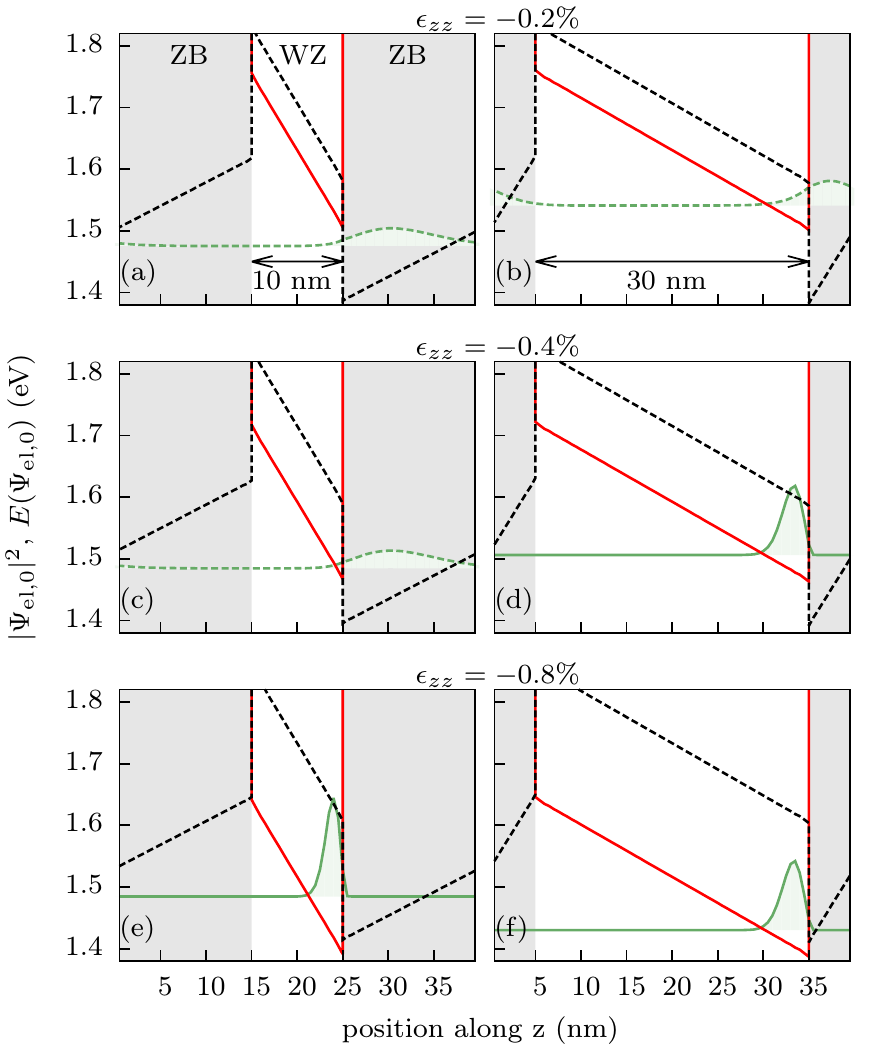}
\caption{(Color online) Charge density and energy of the electron ground state (green) for a 10 [(a), (c), (e)] and 30~nm [(b), (d), (f)] long WZ segment in a periodic superlattice of 40~nm period length for external strains of $\epsilon_{zz} = -0.2$ [(a), (b)], $-0.4$ [(c), (d)], and $-0.8\%$ [(e), (f)]. Black dashed and red solid lines indicate the $\Gamma_{6c,7c}$ and the $\Gamma_{8c}$ bands, respectively. Dashed green lines indicate the electron state to be of $\Gamma_{6c,7c}$ character whereas solid green lines indicate a $\Gamma_{8c}$ character. ZB segments are depicted by shaded gray areas. The plot shows the whole supercell of the simulation.}
\label{fig:rho}
\end{figure}

\subsection{Influence of external strain}
We next study the influence of an additional uniaxial strain $\epsilon_{zz}$ on the electronic properties of WZ/ZB GaAs superlattices. We focus here on the case of a superlattice with $\epsilon_{zz}<0$, for which the interplay of spatially direct and indirect transitions (cf.\ Fig.~\ref{fig:reco}) is most clearly seen.
Figure~\ref{fig:SL}(a) shows the energy difference between electron and hole ground states as a function of the length of the WZ (ZB) segment for different values of $\epsilon_{zz}$. The intrinsic biaxial strain due to the lattice mismatch as well as spontaneous and piezoelectric polarization are taken into account. Upon the application of the external uniaxial strain, the character of the electron ground state changes to $\Gamma_{8c}$ already for shorter WZ segments (for example, 30~nm at $\epsilon_{zz}=-0.2\%$, 15~nm at $-0.6\%$, 5~nm at $-1\%$), as compared to the previously discussed case where the external strain was absent (cf.\ Fig.~\ref{fig:contribs}). 

This change of character can also be seen when examining the charge carrier overlap $\mathcal{O}$ between the electron and hole ground state as defined in Ref.~\onlinecite{MaHa13}. The overlap is below $10^{-5}$ between $\Gamma_{8c}$ electrons and $\Gamma_{9v}$ hole states for WZ segments longer than 10~nm despite the fact that both particles are confined within the same segment implying spatially direct transitions as schematically depicted in Fig.~\ref{fig:reco}. The origin of this unexpected behavior is the polarization potential, which results in a strong confiment of electrons and holes at the opposite facets of the WZ segment. In contrast, the overlap between $\Gamma_{6c}$ electrons confined in the ZB segment and $\Gamma_{9v}$ holes in the WZ segment is much larger ($10^{-4}$ to $10^{-1}$) thanks to the weak confinement of the light $\Gamma_{6c}$ electrons. However, for $\epsilon_{zz} \le -0.8\%$ and short WZ segments, $\mathcal{O}$ increases drastically for $\Gamma_{8c}$ electrons. In these cases, strain reduces the $\Gamma_{8c}$ CB energy to such an amount that the electron remains confined in the WZ segment even for very short segments. 

To illustrate this behavior, Fig.~\ref{fig:rho} shows the charge density of the electron ground state together with the potentials formed by the $\Gamma_{6c,7c}$ and the $\Gamma_{8c}$ bands for WZ segments of 10~nm [cf. Figs.~\ref{fig:rho}(a), (c), (e)] and 30~nm [cf. Figs.~\ref{fig:rho}(b), (d), (f)] length and different values of $\epsilon_{zz}$. For $\epsilon_{zz} = -0.2\%$, $\Psi_\mathrm{el}$ is confined in the ZB segment for both cases [cf. Figs.~\ref{fig:rho}(a) and \ref{fig:rho}(b)], but the wavefunction penetrates into the WZ segment and the confinement of the electron is rather weak. For a strain of $-0.4\%$, the electron remains weakly confined in the ZB segment for a WZ length of 10~nm, but is transfered to the WZ segment and thus changes its character to $\Gamma_{8c}$ for a WZ length of 30~nm [cf. Figs.~\ref{fig:rho}(c) and \ref{fig:rho}(d)]. Evidently, the confinement of the electron in the WZ segment is much stronger due to the large effective mass of the $\Gamma_{8c}$ band along the $\langle$0001$\rangle$ direction, so that tunneling into the ZB segment is negligible. Finally, for a strain of $-0.8\%$, the electron is strongly confined within the WZ segment for both the short and long WZ segment [cf. Figs.~\ref{fig:rho}(e) and \ref{fig:rho}(f)].


\section{Summary and conclusions}
Our findings show that the description of the electronic properties of GaAs polytype nanostructures requires the explicit consideration of the two energetically lowest CBs. We find that the intrinsic strain $\varepsilon_{ij}$ that arises from the lattice mismatch between the two polytypes as well as the piezoelectric and spontaneous polarization have a significant influence on the electronic properties of WZ/ZB GaAs heterostructures and must not be neglected. In particular, both the character of the electron ground state and its energy depend sensitively on the polytype fraction in a given NW. These properties are furthermore affected by external, uniaxial strain acting on the NW. For a range of $-1\% < \epsilon_{zz} < 1\%$, the energy difference between the two relevant CBs of the WZ phase varies between $-250$ and $+200$~meV. The significant influence of comparatively small strains on the optical properties of polytype GaAs NWs is a possible explanation for the controversial experimental results regarding the character of the lowest CB and the energy of the corresponding band gap that were reported in the past. We finally note that many of the parameters employed for our simulations are not known with high accuracy. However, as long as the energy difference between the $\Gamma_{7c}$ and the $\Gamma_{8c}$ CB of the WZ segment is small (as is the case not only in GaAs, but also in GaSb~\cite{BePa12}), the character of the electron ground state will depend on strain state and dimensions of the WZ/ZB heterostructure such that the explicit treatment of the two CBs is required for any simulation of its electronic properties.

\begin{acknowledgments}
The authors thank Friedhelm Bechstedt for his help and valuable suggestions and Lutz Schrottke for a critical reading of the manuscript. P.\,C. acknowledges funding from the Fonds National Suisse de la Recherche Scientifique through project 161032.
\end{acknowledgments}

\section*{Appendix}
The Hamiltonian employed is based on an eight band model by Chuang and Chang,~\cite{ChCh96} where the additional $\Gamma_8$ CB is added:
\begin{widetext}
\begin{equation*}\label{ham}\hat{H}^{10\times 10} =\left(\begin{array}{cccccccccc}
C & 0 & 0 & 0 & R & 0 & 0 & 0 & 0 & 0\\
0 & C & 0 & 0 & 0 & 0 & 0 & 0 & 0 & R\\
0 & 0 & S & 0 & -V & U & V^* & 0 & 0 & 0\\
0 & 0 & 0 & S & 0 & 0 & 0 & -V & U & V^*\\
R & 0 & -V^* & 0 & F & -M^* & -K^* & 0 & 0 & 0\\
0 & 0 & U & 0 & -M & \lambda & M^* & \Delta & 0 & 0\\
0 & 0 & V & 0 & -K & M & G & 0 & \Delta & 0\\
0 & 0 & 0 & -V^* & 0 & \Delta & 0 & G & -M^* & -K^*\\
0 & 0 & 0 & U & 0 & 0 & \Delta & -M & \lambda & M^*\\
0 & R & 0 & V & 0 & 0 & 0 & -K & M & F
\end{array}\right).
\end{equation*}
\end{widetext}

The entities within the matrices are the operators:
\begin{eqnarray*}\label{kppar}
S = E_\mathrm{cb} + A_1'\partial_z^2 + A_2'\left(\partial_x^2 + \partial_y^2\right),\\
F = \Delta_1 + \Delta_2 + \lambda + \theta,\vspace{.1cm}
~~~G = \Delta_1 - \Delta_2 + \lambda + \theta,\\
 \lambda = \frac{\hbar^2}{2m_0}\left(\tilde{A}_1\partial_z^2 + \tilde{A}_2
\left[\partial_ x^2 + \partial_y^2\right]\right) + E_\mathrm{vb},\vspace{.1cm}
\end{eqnarray*}
\begin{eqnarray*}\label{kppar2}
\theta = \frac{\hbar^2}{2m_0}\left(\tilde{A}_3\partial_z^2 + \tilde{A}_4\left[\partial_ x^2 + \partial_y^2\right]\right),\\
K = \frac{\hbar^2}{2m_0}\tilde{A}_5\left(\partial_x + i\partial_y\right)^2,\vspace{.1cm}~~~
M = \frac{\hbar^2}{2m_0}\tilde{A}_6\partial_z\left(\partial_x + i\partial_y\right),\\
 U = i\partial_z P_1,\vspace{.1cm}~~~
V = i\left(\partial_x + i\partial y\right) P_2,~~~
\Delta = \sqrt{2}\Delta_3,
\end{eqnarray*}
with
\begin{eqnarray*}\label{modA}
A_1' = \frac{\hbar^2}{2m_e^\parallel} - \frac{P_1^2}{E_G},~~~A_2' = \frac{\hbar^2}{2m_e^\perp} - \frac{P_2^2}{E_G},\\
\tilde{A}_1 = A_1 + \frac{2 m_0}{\hbar^2}\frac{P_2^2}{E_G},~~~\tilde{A}_2 = A_2\\
\tilde{A}_3  =  A_3 - \frac{2 m_0}{\hbar^2}\frac{P_2^2}{E_G},~~~\tilde{A}_4 = A_4 + \frac{2 m_0}{\hbar^2}\frac{P_1^2}{E_G},
\end{eqnarray*}
\begin{eqnarray*}\label{modA2}
\tilde{A}_5 = A_5 + \frac{2 m_0}{\hbar^2}\frac{P_1^2}{E_G},~~~\tilde{A}_6 = A_6 + \frac{\sqrt{2} m_0}{\hbar^2}\frac{P_1 P_2}{E_G},
\end{eqnarray*}
\begin{eqnarray*}\label{PDelta}
P_1^2  = \frac{\hbar^2}{2m_0}\left(\frac{m_0}{m_e^\perp}-1\right)\\
\times\frac{(E_G + \Delta_1 + \Delta_2)(E_G + 2\Delta_2)-2\Delta_3^2}{E_G + 2\Delta_2},\\
P_2^2  = \frac{\hbar^2}{2m_0}\left(\frac{m_0}{m_e^\parallel}-1\right)\\
\times\frac{E_G[(E_G + \Delta_1 + \Delta_2)(E_G + 2\Delta_2)-2\Delta_3^2]}{(E_G + \Delta_1 + \Delta_2)(E_G + \Delta_2) - \Delta_3^2},\\
\Delta_1 = \Delta_\mathrm{cr}~~~\mathrm{and}~~~\Delta_2=\Delta_3=\frac{1}{3}\Delta_\mathrm{so}.
\end{eqnarray*}
$E_\mathrm{cb}$ and $E_\mathrm{vb}$ denote the conduction and valence band edge,
$E_G=E_\mathrm{cb}-E_\mathrm{vb}$ is the band gap, and $m_0$ is
the bare electron mass. $m^\parallel_e$ and $m^\perp_e$ are the electron effective masses of the $\Gamma_6$ (ZB) and $\Gamma_7$ (WZ) CB,
$\Delta_\mathrm{cr}$ and $\Delta_\mathrm{so}$ denote the crystal field and the spin-orbit
splitting parameter, respectively. $A_1$ to $A_6$ are the Luttinger-like parameters.
The $\Gamma_8$ band is added via the term:
\begin{equation*}
C = E_\mathrm{cb} + \Delta E(\Gamma_8,\Gamma_7) + \frac{\hbar^2}{2m_{8,\parallel}}\partial_z^2 + \frac{\hbar^2}{2m_{8,\perp}}(\partial_x^2+\partial_y^2)
\end{equation*}
Here, $\Delta E(\Gamma_8, \Gamma_7)$ denotes the energy splitting between the two bands at the $\Gamma$-point and $m_{8,\parallel}$ and $m_{8,\perp}$ denote
the effective mass along the [0001] direction or perpendicular to it. $R \approx 0$ denotes the small, but dipole-allowed coupling of the $\Gamma_8$ CB and the $\Gamma_{9v}$ VB.
Strain enters the Hamiltonian via the additional contribution:
\begin{equation}
\hat{H}_\mathrm{strain} = \left(\begin{array}{cccccccccc}
c & 0 & 0 & 0 & 0 & 0 & 0 & 0 & 0 & 0\\
0 & c & 0 & 0 & 0 & 0 & 0 & 0 & 0 & 0\\
0 & 0 & s & 0 & 0 & 0 & 0 & 0 & 0 & 0\\
0 & 0 & 0 & s & 0 & 0 & 0 & 0 & 0 & 0\\
0 & 0 & 0 & 0 & f & -h^\star & - k^\star & 0 & 0 & 0\\
0 & 0 & 0 & 0 & -h & \lambda_\epsilon & h^\star & 0 & 0 & 0\\
0 & 0 & 0 & 0 & -k & h & f & 0 & 0 & 0\\
0 & 0 & 0 & 0 & 0 & 0 & 0 & f & -h^\star & -k^\star\\
0 & 0 & 0 & 0 & 0 & 0 & 0 & -h & \lambda_\epsilon & h^\star\\
0 & 0 & 0 & 0 & 0 & 0 & 0 & -k & h & f
\end{array}\right)
\end{equation}
where:
\begin{equation}\label{wzParamStrain}
\begin{array}{l}
\displaystyle
c = (\Xi_{d,h}-\Xi_{b,h})\cdot \tau \cdot \epsilon_{zz} + \Xi_{d,u}\cdot (1-\tau)\cdot\epsilon_{zz},\vspace{.1cm}\\
s = (\Xi_{b,h}-D_1-2D_2)\cdot \tau\cdot\epsilon_{zz} + D_3\cdot (1-\tau)\cdot \epsilon_{zz},\vspace{.1cm}\\
\tau = (1-2\nu)/3~~\mathrm{and}~~\nu= C_{12}/(C_{12}+C_{11}),\vspace{.1cm}\\
\displaystyle
\lambda_\epsilon = D_1\epsilon_{zz} + D_2(\epsilon_{xx}
      + \epsilon_{yy}),\vspace{.1cm}\\
\displaystyle
\theta_\epsilon = D_3\epsilon_{zz} + D_4(\epsilon_{xx}
      + \epsilon_{yy}),\vspace{.1cm}\\
\displaystyle
f = \lambda_\epsilon + \theta_\epsilon,\vspace{.1cm}\\
\displaystyle
k = D_5(\epsilon_{xx} + 2i\epsilon_{xy} - \epsilon_{yy}),\vspace{.1cm}\\
\displaystyle
h = D_6(\epsilon_{zx} + \rm{i} \epsilon_{yz}).
\end{array}
\end{equation}

\end{document}